\documentclass{article}

\usepackage{arxiv}

\usepackage[utf8]{inputenc} 
\usepackage[T1]{fontenc}    
\usepackage{hyperref}       
\usepackage{url}            
\usepackage{booktabs}       
\usepackage{amsfonts}       
\usepackage{nicefrac}       
\usepackage{microtype}      
\usepackage{lipsum}		
\usepackage{graphicx}
\usepackage{doi}
\usepackage{float}
\usepackage{subfigure}

\title{Comparison  of computing efficiency among FFT, CZT and Zoom FFT in THz-TDS}

\author{ 
	{\hspace{1mm}Abel García-Devesa, Miguel A. Báez-Chorro, Borja Vidal}\\
	Nanophotonics Technology Center\\
	Universitat Politècnica de València\\
	C/Camí de Vera, sn, 46022, Valencia,Spain \\
	\texttt{abgarde@teleco.upv.es, mibaecho@ntc.upv.es, bvidal@dcom.upv.es} \\
}

\renewcommand{\headeright}{\thepage}

\hypersetup{
}

\begin{document}
\maketitle

\begin{abstract}
	A study of alternative transforms to FFT in order to compare their potential to enhance resolution and computation time in the framework of THz time domain spectroscopy (THz-TDS) instruments is carried out. Both from simulated and experimental data it is shown that, as expected, resolution cannot be enhanced using CZT or Zoom FFT and, in terms of computing efficiency, FFT is in practical cases, the most efficient one.
\end{abstract}


\section{Introduction}
The Terahertz (THz) band is a valuable asset to extract information from a wide range of materials and industrial processes [1-3]. THz radiation can penetrate nonpolar dielectrics, as well as microwaves, but it has better resolution due to its shorter wavelength and, being non-ionizing, these instruments are easier to operate than X-rays.
THz sensing is usually performed using THz Time Domain Spectroscopy (THz-TDS). THz-TDS instruments, especially when implemented using optical fiber [4], are very attractive for industrial quality control since they can be easily deployed with a compact sensing head that is connected through a robust fiber link to a remote unit [5]. \\

These systems are usually based on acquiring the electric field of the THz wave using photoconductive antennas [6] and deriving the amplitude and phase of the waves to extract the optical constants of the material under test [7]. Another approach is based on ellipsometry [8]. In any of these systems, information is obtained in the time domain. Material information is extracted from the frequency response of the sample which is derived by comparing the information of the sample trace to the reference one. \\

Spectral information is conventionally derived from the time domain measurements using the discrete Fourier transform (DFT) [1-11].
Here, alternative methods to obtain spectral estimates in the framework of THz-TDS are reviewed and compared. Tests are carried out to show that they do not provide enhanced resolution. Then, the computation efficiency is compared, showing that conventional FFT is the most efficient approach.

\section{Theory}
\label{sec:headings}
\subsection{Generalized FFT}
In THz-TDS, the photocurrent at the output of the receiving photoconductive antenna  is sampled to obtain a sequence of points which define the THz trace to be processed.
\begin{equation}
x(t)= x(tT_s)= x[n] \hspace{0.5cm} \forall \hspace{0.5cm} T_s= \frac{1}{f_s}
\end{equation}
where $T_s$ s the sampling period and $f_s$
is the sampling frequency. \\

The DFT of a time signal can be defined as \cite{b12}:
\begin{equation}
X[k]= \sum_{n=0}^{N-1} x[n] W^{-kn} \hspace{0.5cm} \forall \hspace{0.5cm} k=0,...,N-1
\label{eq_DFT} 
\end{equation}

where $W=e^{-j\frac{2 \pi }{N}}$ and N is the number of samples. The DFT requires  $O(N^2)$ arithmetic operations. The fast Fourier transform, known as the FFT, is a more efficient algorithm that allows the discrete Fourier transform to be computed in $O(N_{FFT}log_2(N_{FFT}))$ operations \cite{b13}.\\

\begin{figure}[htbp]
\centering
\includegraphics[width=5cm]{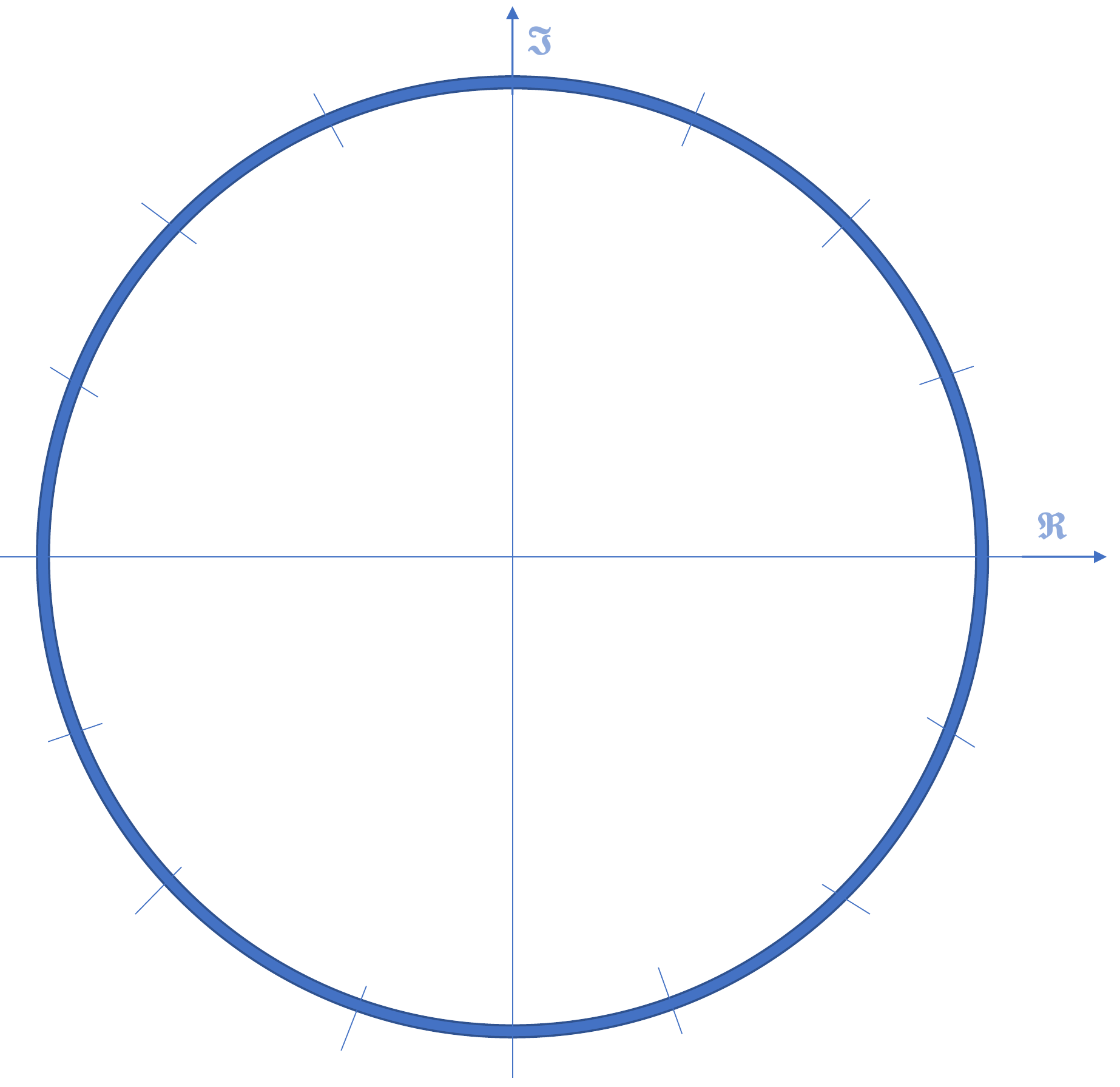}
\caption{FFT distribution in the Z-plane.}
\label{imag_fftz} 
\end{figure}

In the z-plane, the DFT is represented as a circle of radius unity with the points equiespacied, as seen in Figure \ref{imag_fftz}.

The resolution of the FFT is:
\begin{equation}
\Delta f= \frac{f_s}{N_{FFT}}
\label{eq_resfft}
\end{equation}

From (\ref{eq_resfft}), it can be seen that the spectral resolution depends on the number of samples. It is widely known that, by adding zeros at the end of the time signal, the plotting resolution of the spectral representation is improved. This process, known as zero padding, does not provide additional information and thus, the spectral resolution is not enhanced. The effect of zero padding can be seen just as an interpolation of the data, which improves plotting resolution.


In the case of looking for a specific resolution, the number of zeros to be added can be determined by increasing the plotting resolution  in frequency ($\Delta$). Considering (\ref{eq_resfft}), a difference in resolution is defined by adding x zeros:
\begin{equation}
\Delta=\Delta f - \Delta f' =\frac{f_s}{N_{FFT}} - \frac{f_s}{N_{FFT}+x}
\label{eq_desarrollo}
\end{equation}

From (\ref{eq_desarrollo}) the number of zeros to be added is:
\begin{equation}
x=\frac{f_s L}{f_s - \Delta \cdot N_{FFT}} -N_{FFT}
\end{equation}

The question now is: is it possible to use a different mathematical tool to enhance resolution or to enhance computation time? In the literature, two methods claim to improve resolution or computation time: Zoom FFT and CZT. Next, these are briefly reviewed.
\subsection{Zoom FFT}
When optimising the FFT, a kind of frequency zoom can be used to avoid having to analyse the whole spectrum. This method is known as Zoom FFT  \cite{b14}. It is used to obtain a given plotting resolution in a band while reducing the number of samples through a series of operations prior to performing the FFT. The scheme used is as follows:

\begin{figure}[H]
\centering
\includegraphics[width=14cm]{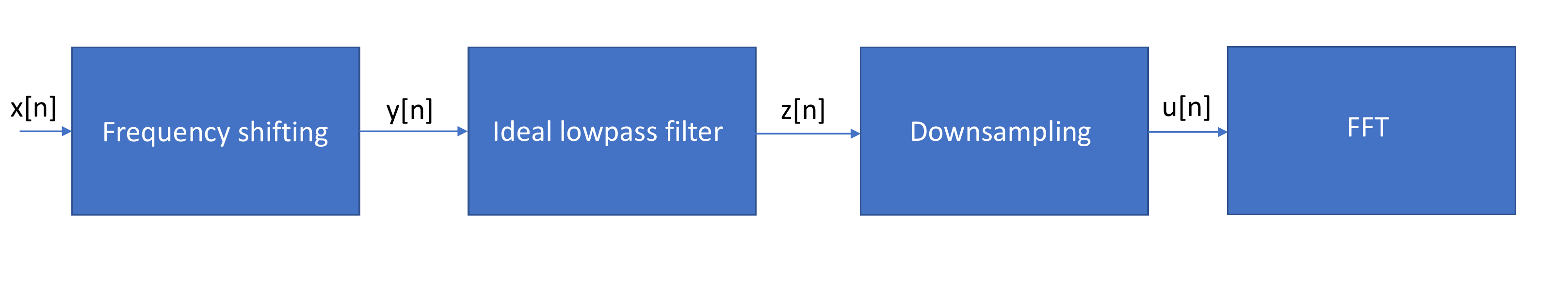}
\caption{Zoom FFT block diagram.}
\label{fig_diag_zoomfft} 
\end{figure}

The idea of Zoom FFT is to choose a region of the spectrum between $f_1$ and $f_2$ and perform only the FFT of this band. The steps to be performed, which can be seen in Figure \ref{fig_diag_zoomfft}, are as follows:

\begin{itemize}
\item Frequency shifting: since it is desired to concentrate the signal between $f_1$ and $f_2$, the spectrum is centred at:
\begin{equation}
f_N=\frac{f_1+f_2}{2}
\end{equation}
For this purpose, a frequency shift is performed in the time domain:
\begin{equation}
y[n]=x[n]e^{jN_{FFT}f_N}
\end{equation}
\item Ideal lowpass filter: lowpass filtering is performed:
\begin{equation}
H[k]= \left\{ \begin{array}{lcc}
             C &   if  & -f_c \leq k \leq f_c \\
             \\ 0 &  otherwise \\
             \end{array}
   \right.  
\end{equation}

\item Downsampling: savings come from being able to calculate a much shorter FFT while achieving the same resolution. For a decimation factor of D, the new sampling frequency is:
\begin{equation}
f_{sD}=\frac{f_s}{D}
\end{equation} 
So the length of the spectrum is $L_D = \frac{L}{D}$. The operation to be performed in the time domain is:
\begin{equation}
u[m]=z[mD]=\sum_{k=0}^{N-1}h[k]y[mD-k]
\end{equation}
\end{itemize}

The computational cost of Zoom FFT is $O(\frac{N_{FFT}}{2D}log_2(\frac{N_{FFT}}{D}))$ \cite{b14}, but a number of extra operations are needed to shift, filter and downsample the signal.

\subsection{Chirp-Z Transform}
The Chirp-Z Transform (CZT) is a generalization of the DFT \cite{b15}. It is defined as follows:
\begin{equation}
X[k]=\sum_{n=0}^{N-1} x[n] A^{-n} W^{kn}    \hspace{0.5cm} \forall \hspace{0.5cm} k=0,...,N_{CZT}-1
\label{eq_CZT}
\end{equation}

where A is the complex starting point (initial frequency in Z-plane), W is the complex relation between the points and M is the number of points of the CZT. If $A=1$,$W=e^{-j\frac{ 2 \pi}{N_{CZT}}}$ and $N_{CZT}=N_{FFT}$,
it gives equation (\ref{eq_DFT}), which corresponds to the DFT. The CZT concept is based on expanding the DFT terms as follows:

\begin{equation}
kn=\frac{k^2+n^2}{2}-\frac{(k-n)^2}{2}
\end{equation}
Substituting it into the equation (\ref{eq_CZT}) leaves:
\begin{equation}
X[k]=W^{\frac{m^2}{2}}\sum_{n=0}^{N-1} x[n] A^{-n} W^{\frac{n^2}{2}} W^{-\frac{(m-n)^2}{2}}
\end{equation}

This equation can be considered as a circular convolution or equivalently as performing two FFTs, one IFFT and four complex multiplications:
\begin{equation}
X[k]=W^{\frac{m^2}{2}} \left[ \left( x[n]A^{-n} W^{\frac{n^2}{2}} \right) \ast W^{-\frac{n^2}{2}} \right]
\end{equation}
\begin{equation}
X[k]=W^{\frac{m^2}{2}} IFFT \left[ FFT \left( x[n]A^{-n} W^{\frac{n^2}{2}} \right) FFT \left( W^{-\frac{n^2}{2}} \right) \right]
\end{equation}

Considering the following definitions:
\begin{equation}
A=A_o e^{j \theta_o}
\label{eq_A}
\end{equation}
\begin{equation}
W=W_o e^{-j \phi_o}
\label{eq_W}
\end{equation}
where $A_o$ and $\theta_o$ define the frequency of the initial sample and $\phi_o$ defines the angle increment between samples. In the case where $A_o$ and $W_o$ are one, the transform is calculated on the unit circle but it will be limited to a particular area of the spectrum from $\theta_o$ and $\phi_o$ which will determine the frequency: $f=[f_{min};f_{max}]$.

\begin{figure}[htbp]
\centering
\includegraphics[width=5cm]{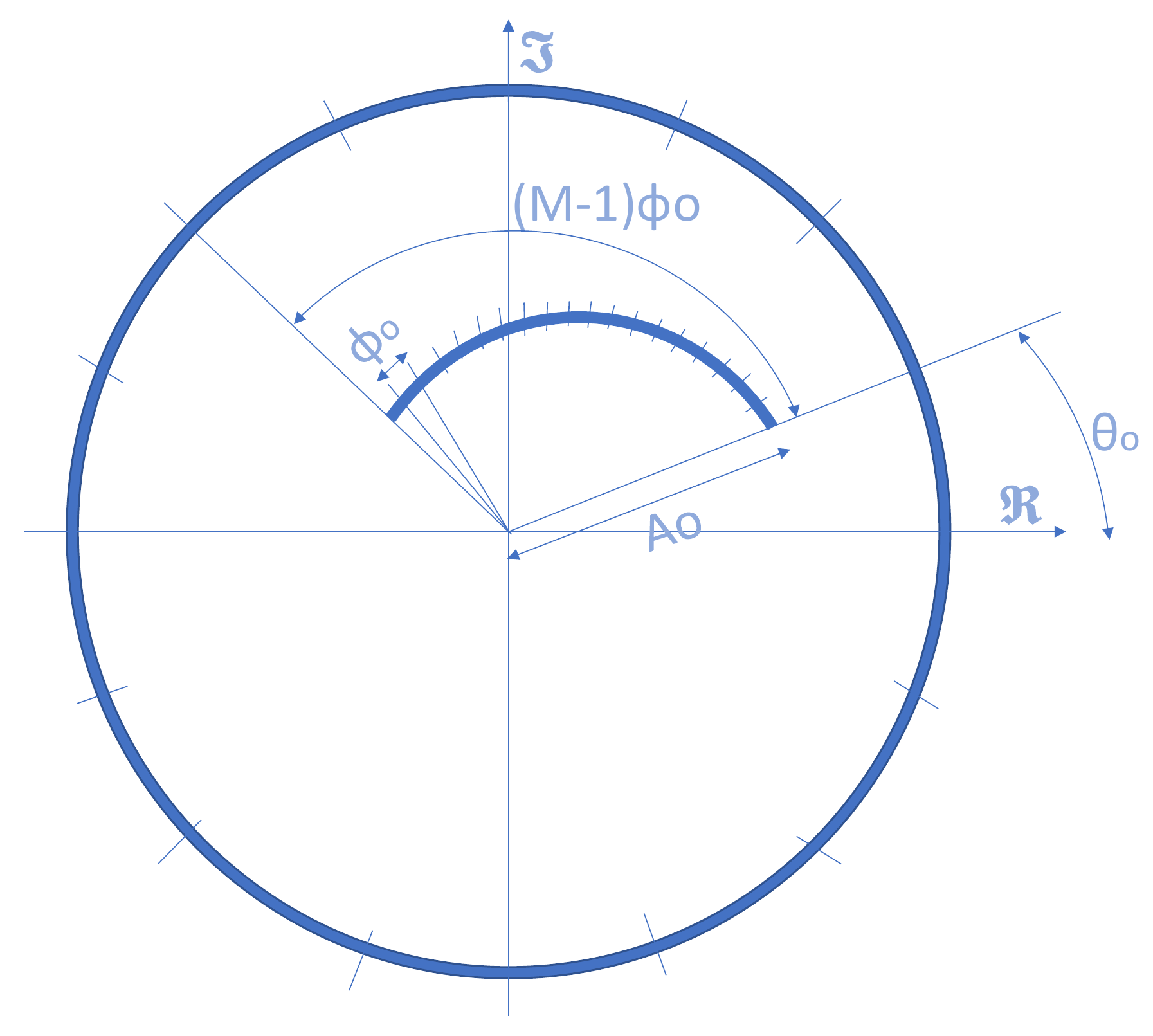}
\caption{Z-plane CZT distribution.} 
\label{fig_CZT}
\end{figure}

The CZT parameters can be obtained by looking at Figure \ref{fig_CZT}, considering the full circle of the Z-plane, so:
\begin{equation}
 \frac{2 \pi}{f_s}=\frac{(N_{CZT}-1) \phi_o}{f_{max}-f_{min}}
 \label{eq_czt1}
 \end{equation}
 \begin{equation}
 \frac{2 \pi}{f_s}=\frac{\theta_o}{f_{min}}
 \label{eq_czt2}
 \end{equation}

From equations (\ref{eq_czt1}) and (\ref{eq_czt2}) it follows that:
\begin{equation}
\theta_o=\frac{2 \pi (f_{max}-f_{min})}{f_s (N_{CZT}-1)}
\end{equation}
\begin{equation}
\phi_o=\frac{2\pi f_{min}}{f_s}
\end{equation}

To obtain the CZT parameters, only the maximum frequency, the minimum frequency and the desired number of points need to be determined. If these values are substituted into (\ref{eq_A}) and (\ref{eq_W}):
\begin{equation}
W=W_o e^{-j \frac{2 \pi (f2-f1)}{N_{CZT} f_s}} 
\end{equation}
\begin{equation}
A=A_o e^{j\frac{2 \pi f1}{f_s}}
\end{equation}

Compared to the FFT, in theory, the frequency resolution will be arbitrary, as the parameters are arbitrary:
\begin{equation}
\Delta f= \frac{f_W}{N_{CZT}}=\frac{f_{max}-f_{min}}{N_{CZT}}
\label{eq_resczt}
\end{equation}

If CZT has to have the same resolution than the FFT, equations (\ref{eq_resfft}) and (\ref{eq_resczt}) must be equated and the number of needed samples can be expressed as:
\begin{equation}
N_{CZT}=\frac{f_2-f_1}{f_s}N_{FFT}
\label{eq_samples}
\end{equation}
In CZT, a greater degree of freedom is achieved when performing the transform, since by choosing the parameters A, W and M, the frequency resolution can be set and an initial frequency and a final frequency can be chosen, as well as the desired number of samples. If it is desired to work on the unit circle, only $W_o$ and $A_o$ with unit value have to be selected. For CZT, the operational cost is $O(Nlog_2(N))$, where N is the maximum of $N_{CZT}$ and $N_{FFT}$ \cite{b16} \cite{b17}.

\section{Performance comparison}

\subsection{Two-tone performance}
\begin{figure}[H]
\centering
\subfigure{\includegraphics[width=7cm]{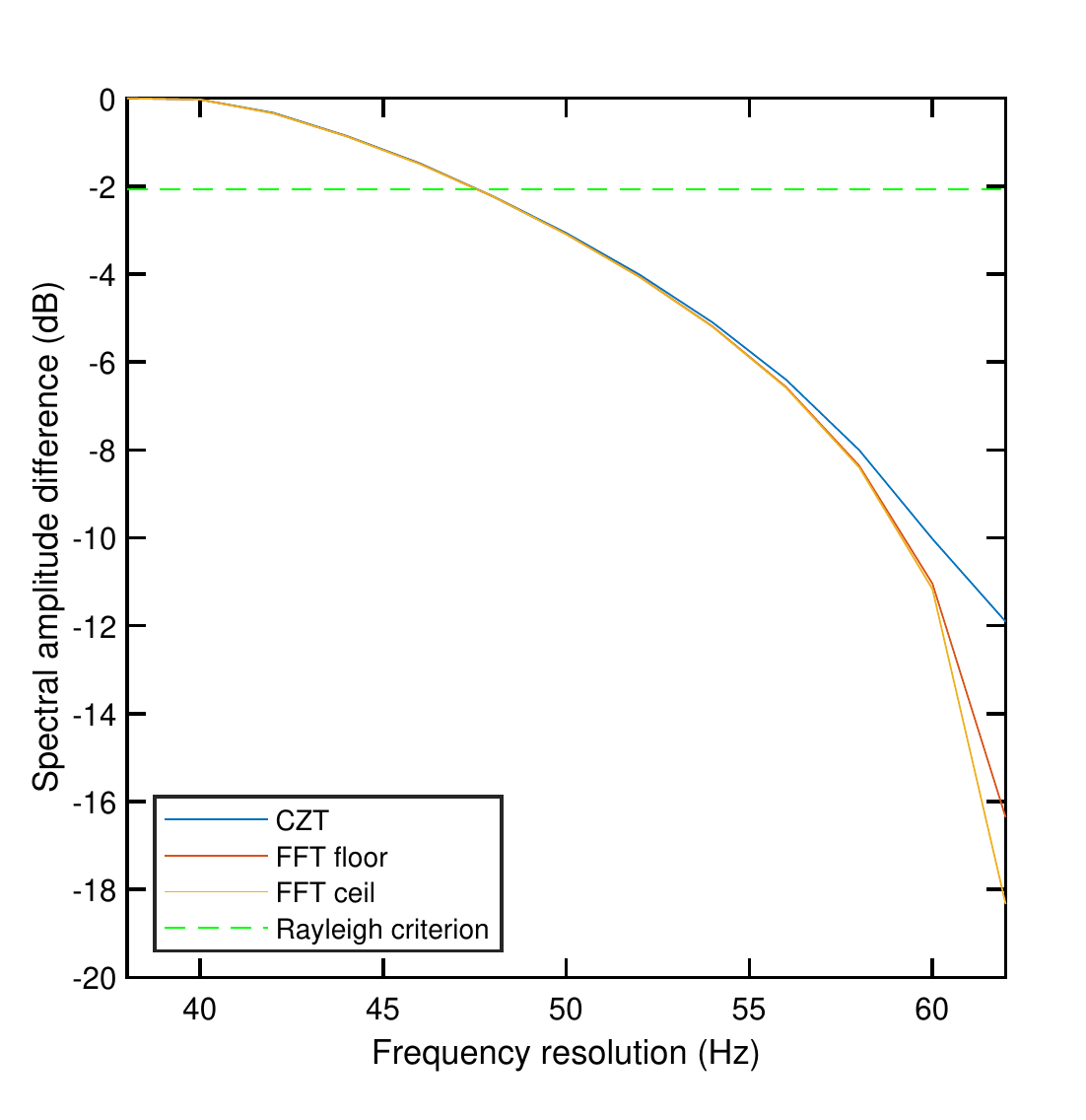}} 
\subfigure{\includegraphics[width=7cm]{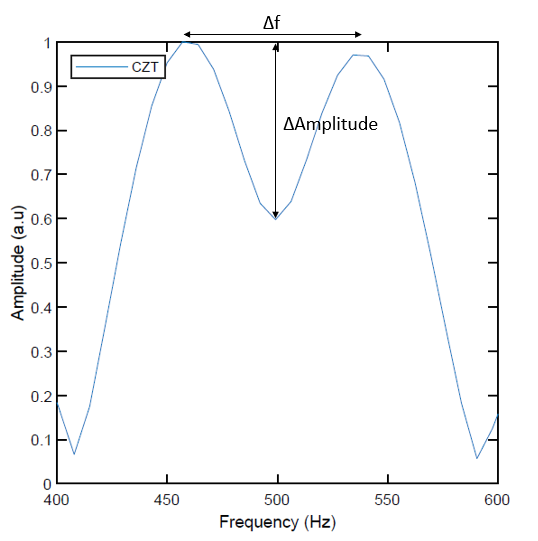}}
\caption{Difference in amplitude between two tones for different difference frequency tones (left). CZT for 475 Hz and 525 Hz (right).}
\label{fig_nevada}
\end{figure}

To test with simple signals, a sine signal is generated as the sum of two sines of different frequencies, with 128 samples and a sampling rate of 8000 Hz. Performing the CZT focusing on the band from $f_1=100$ Hz to $f_2=1000$ Hz gives a plotting resolution of 7.03 Hz. The plotting resolution of the FFT can be approximated to the plotting resolution of the CZT by performing zero padding. The ability to distinguish between the two tones can be quantified for CZT and FFT by measuring the distance between the maximum amplitude closer to the frequency of the tone and the local minimum between them. If a tone distinction criterion is established, such as the Rayleigh criterion, which states that two tones are distinguishable if the null of the first one coincides with the maximum of the first one. It can be seen in Figure \ref{fig_nevada}(left) that a frequency tone cannot be distinguished earlier in the CZT than in the FFT. 

Looking at the Z-plane with respect to the unit circle corresponding to the equispaced FFT, it can be seen that the CZT only uses part of the unit circle, i.e. only part of the spectrum.  
\begin{figure}[H]
	\includegraphics[width=7cm]{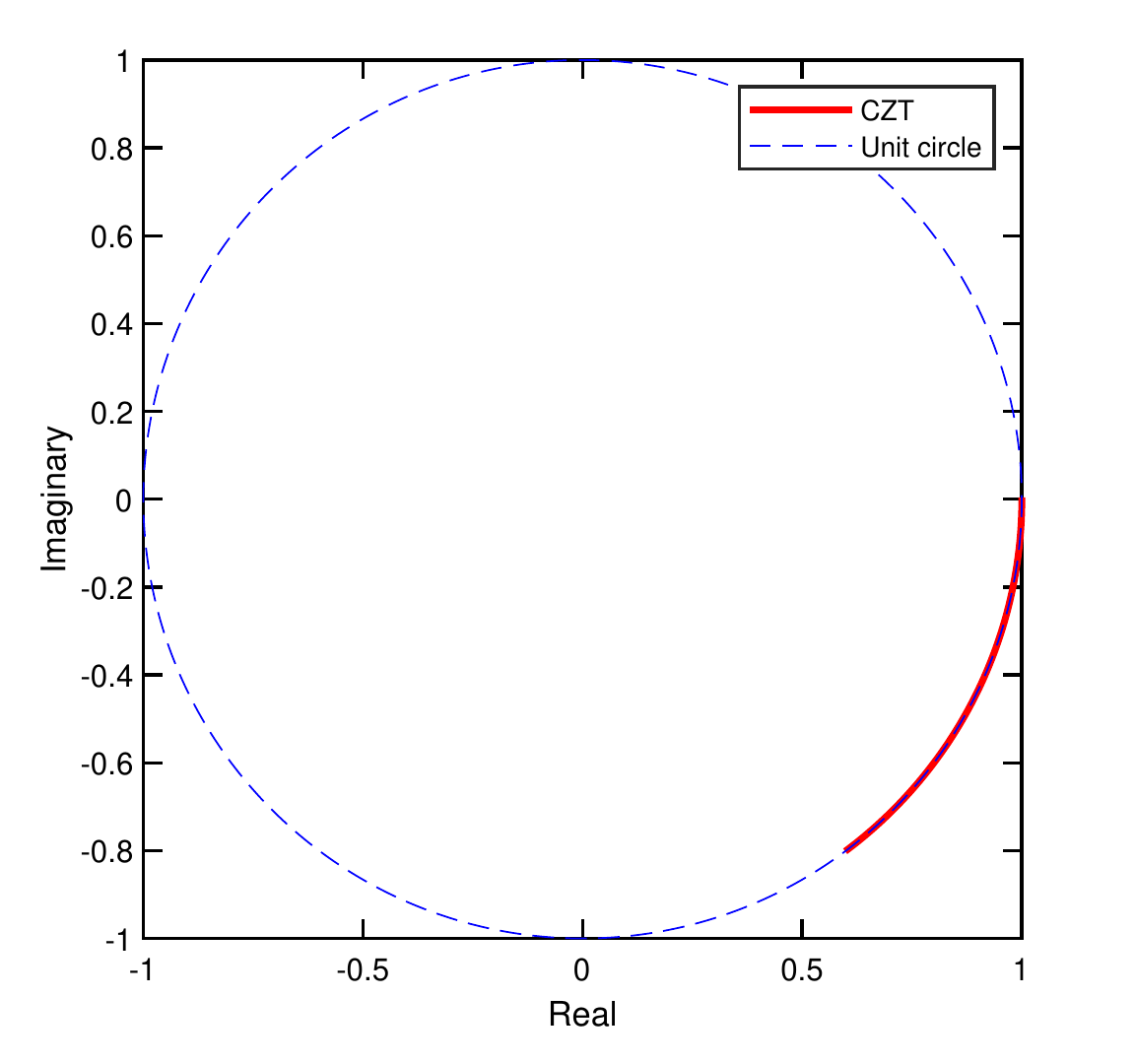}
	\centering
	\caption{Z-Plane of CZT.} 
\end{figure}

\subsection{Perfomance comparison in THz-TDS}

\begin{figure}[htbp]
\hfill
	
	\includegraphics[width=7cm]{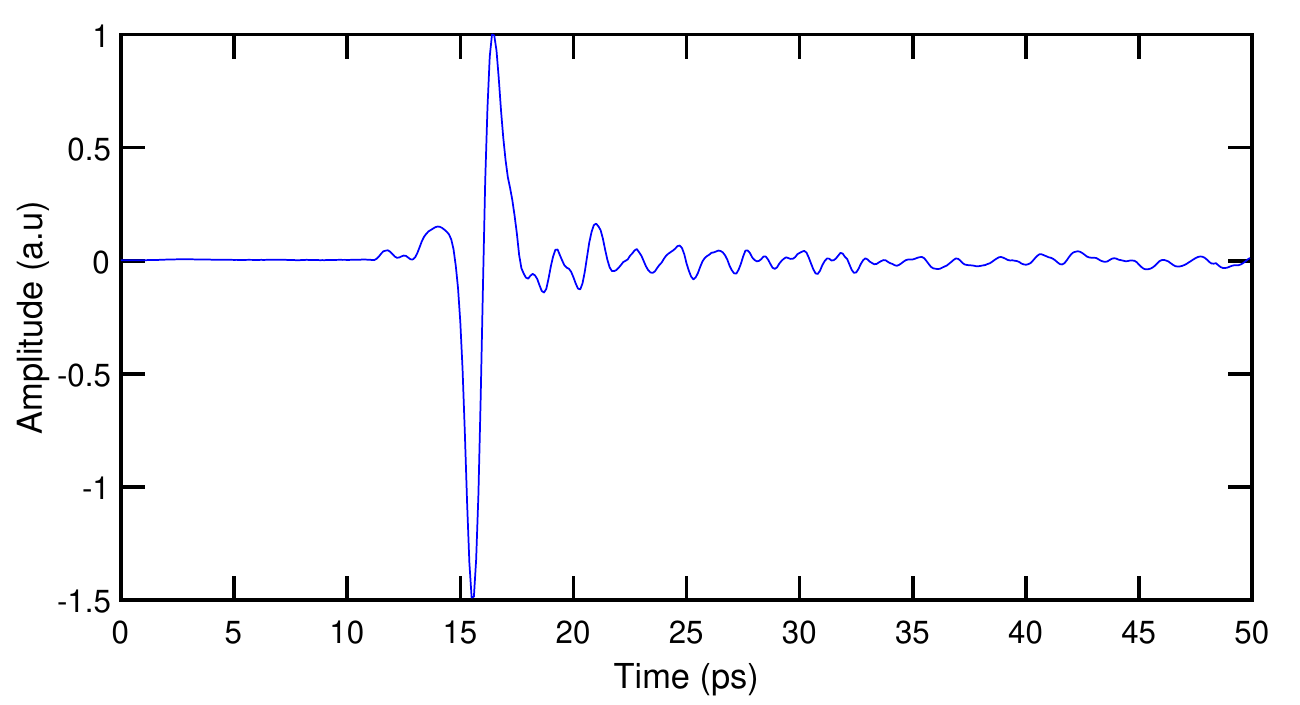}
	\centering
	\caption{Temporary signal of 500 samples.} 
	\label{fig_senyal_lockin1}
\hfill
\end{figure}

\begin{figure}[H]
\centering
\subfigure{\includegraphics[width=7cm]{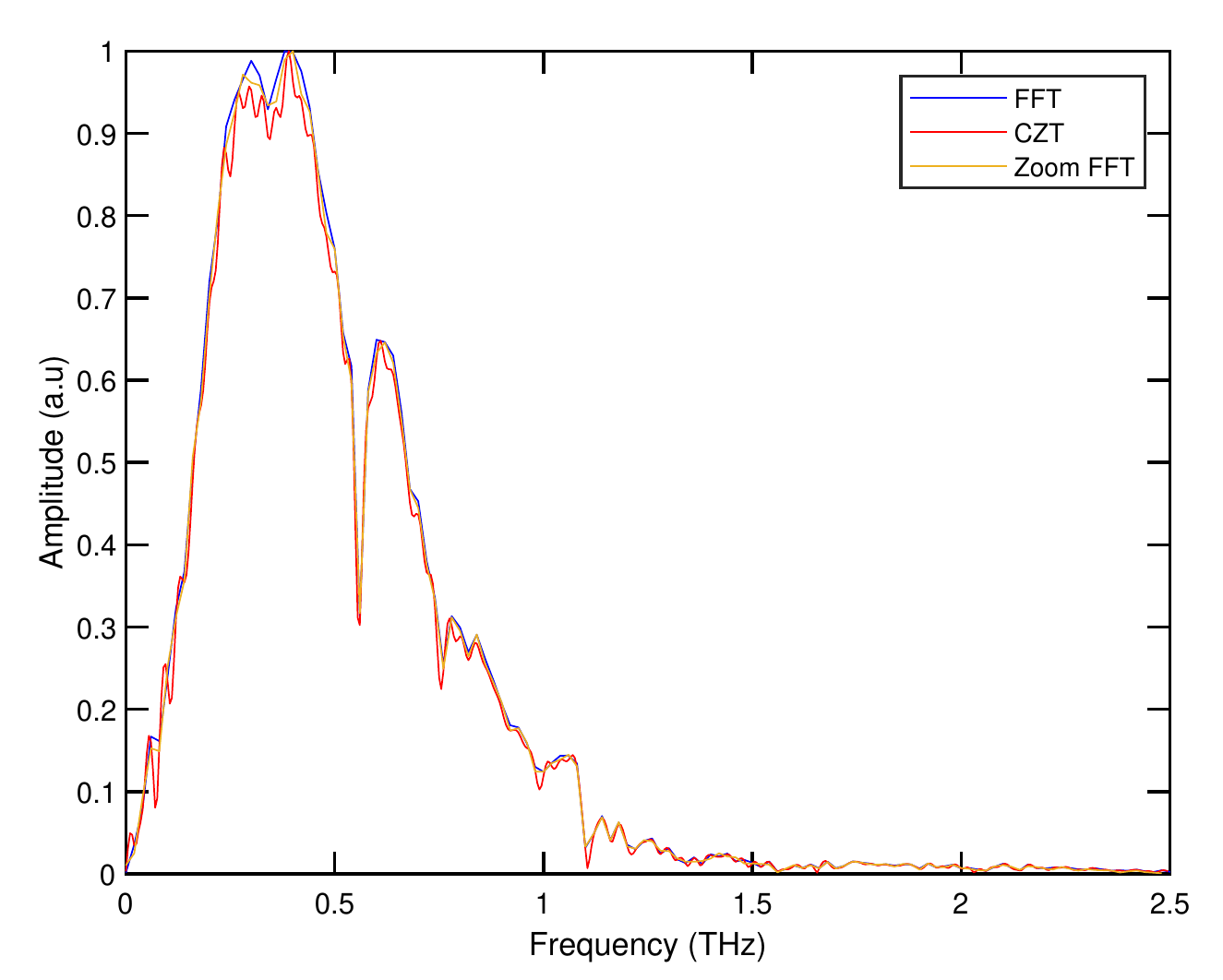}} 
\subfigure{\includegraphics[width=7cm]{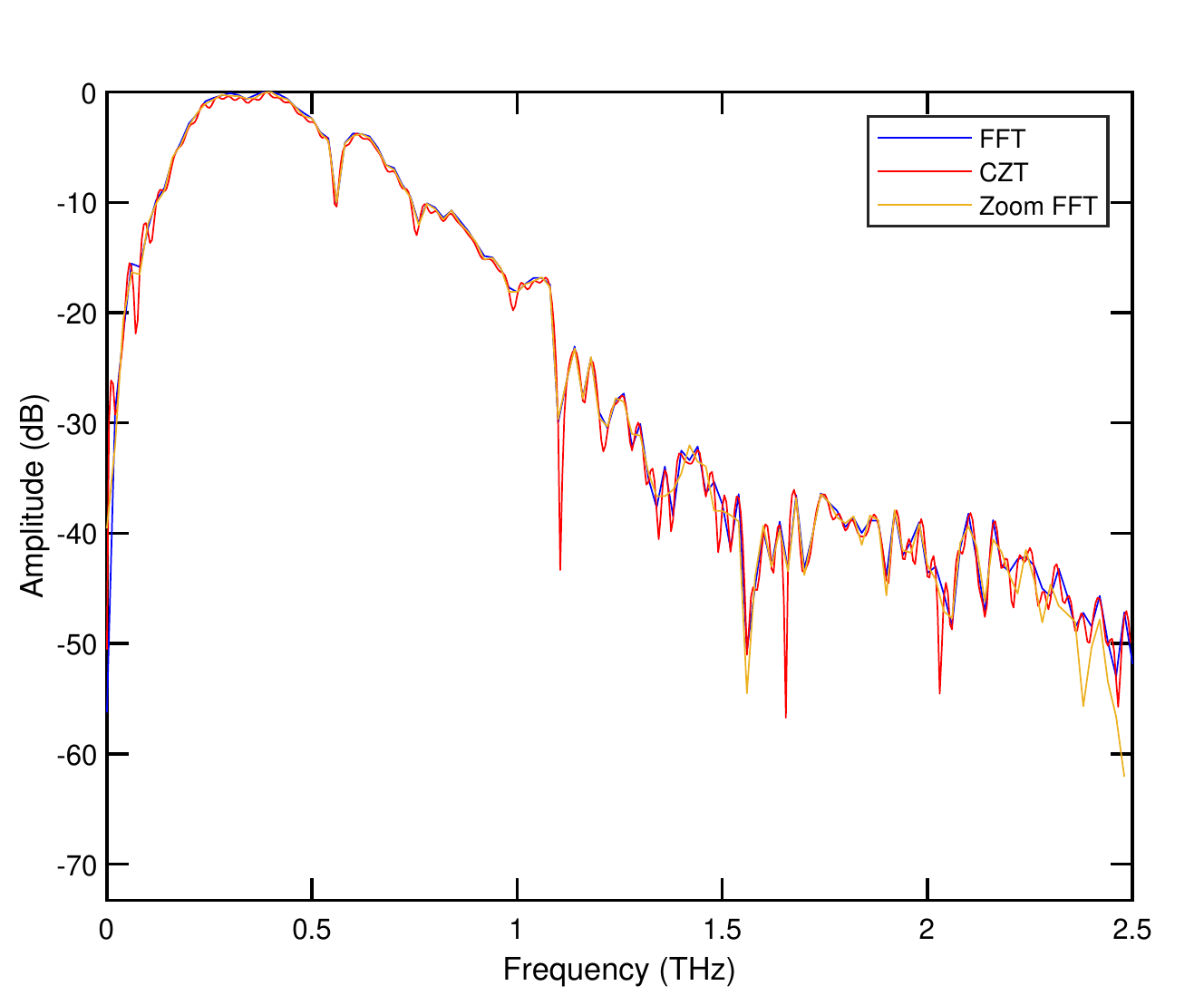}}
\caption{Comparison of FFT, Zoom FFT and CZT of the reference signal from 0 to 2.5 THz:(left) linear; (right) logarithmic.}
\label{fig_comp_1}
\end{figure}

In THz-TDS it is interesting to enhance resolution and computation time. Here, experimental THz-TDS traces have been processed using FFT, Zoom FFT and CZT.
The spectrum of THz-TDS signals has been derived from experimental measurements. A trace can be seen in the Figure \ref{fig_senyal_lockin1}, with $T_s=100$ fs and $N=500$, and their corresponding transformations in Figure \ref{fig_comp_1}.

The plotting resolution of the FFT and the Zoom FFT is $\Delta f=20$ GHz. For the CZT is $\Delta f=2.98$ GHz. Figure \ref{fig_comp_1} shows that to achieve the same plotting resolution, the number of points is lower in the Zoom FFT, since a decimation occurs, and in the CZT, since the number of points is concentrated in the desired area of the spectrum. The lobes observed in the CZT are caused by windowing, as a false frequency resolution is being added and as a consequence the nulls of the sinc, which is the transformation of a square window, are observed.\\

\begin{figure}[H]
\hfill
	
	\includegraphics[width=7cm]{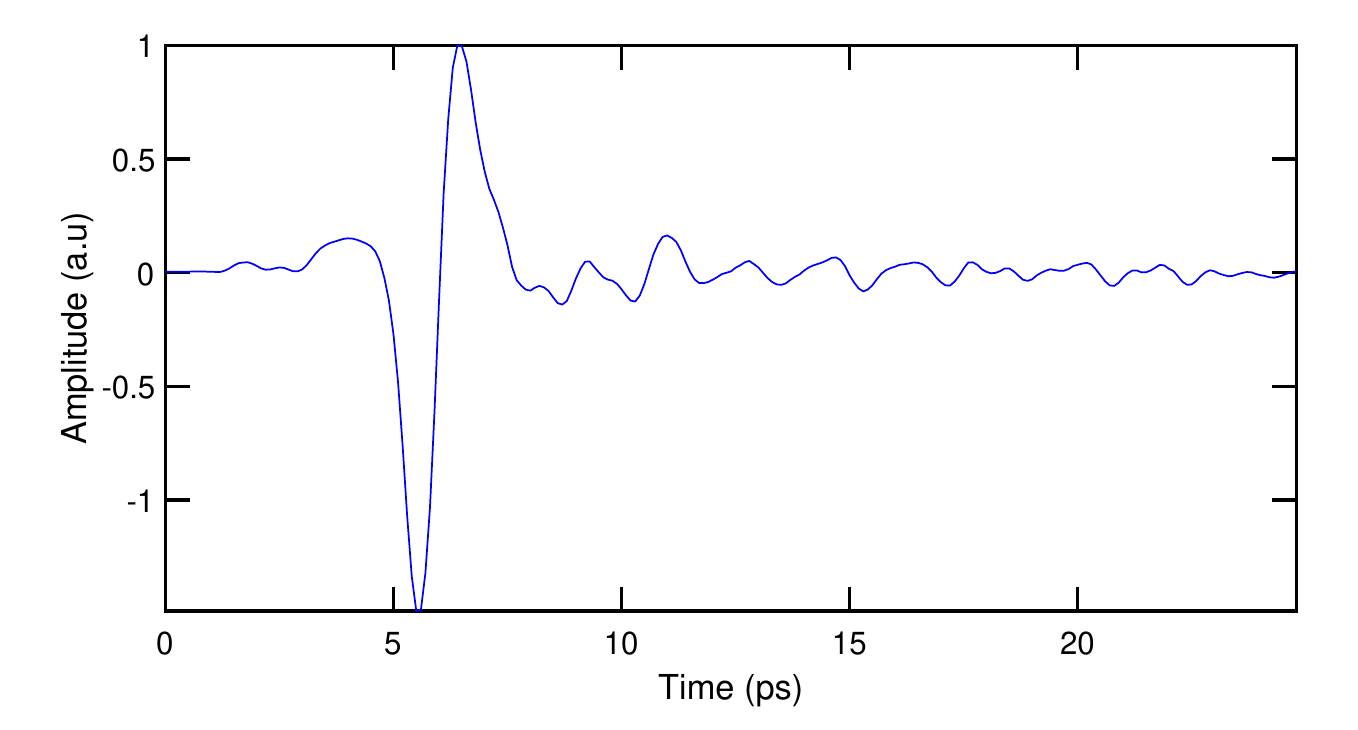}
	\centering
	\caption{Time domain signal of 250 samples.} 
\hfill
\label{fig_senyal_lockin2}
\end{figure}

\begin{figure}[H]
\centering
\includegraphics[width=7cm]{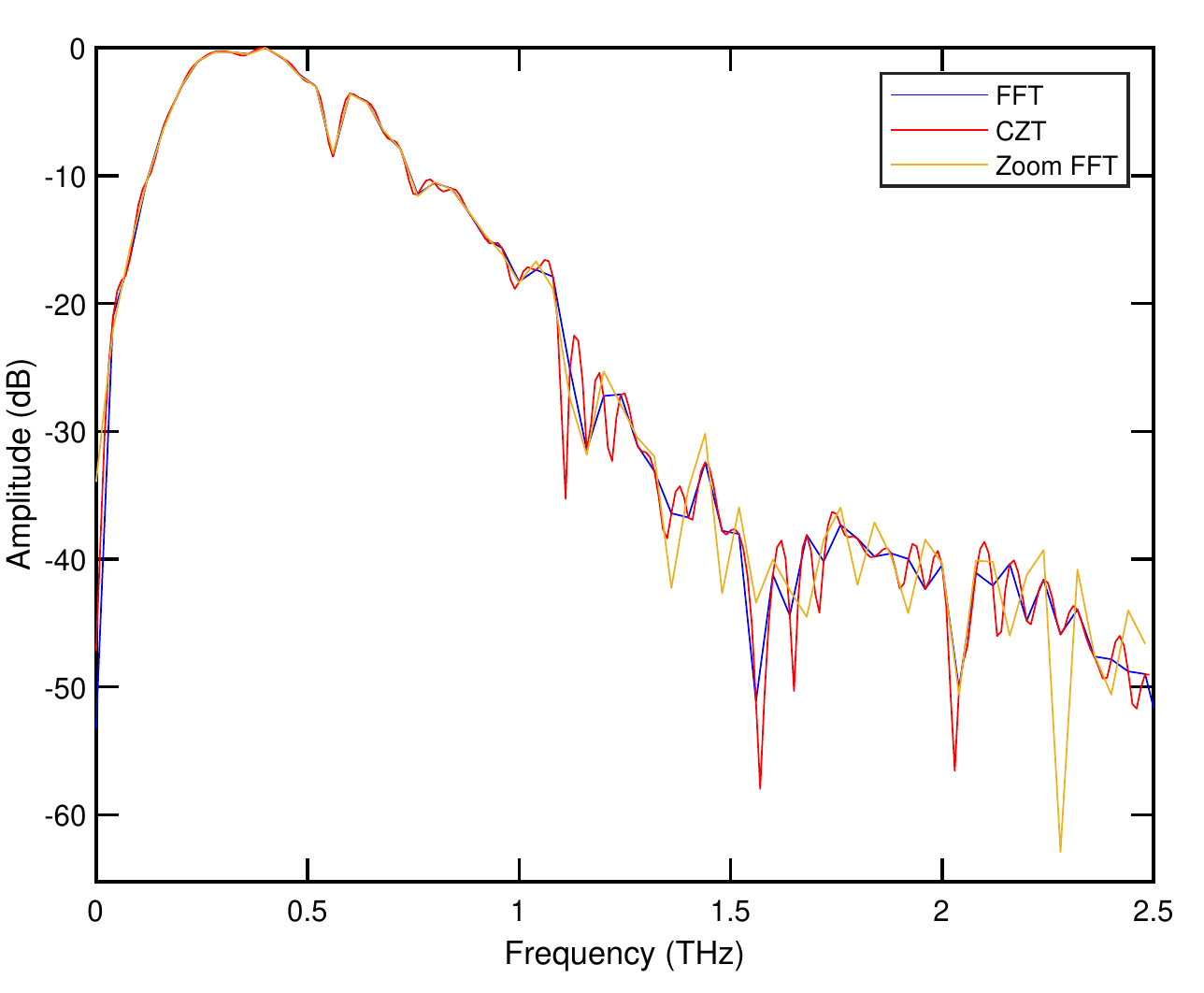}
\caption{Comparison of FFT and CZT of the reference signal from 0 to 2.5 THz}
\label{fig_comp_2}
\end{figure}

By reducing the number of samples to 250 and removing the part of the signal of least interest it can be checked whether a different frequency resolution can be observed with the same window. It is shown in Figure \ref{fig_senyal_lockin2} and \ref{fig_comp_2}.

The resolution of the FFT and Zoom FFT is $\Delta f=40$ GHz while the CZT one is $\Delta f=5.96$ GHz. By reducing the number of samples and increasing the resolution, the signal information is reduced.

\begin{figure}[H]
\centering
\includegraphics[width=7cm]{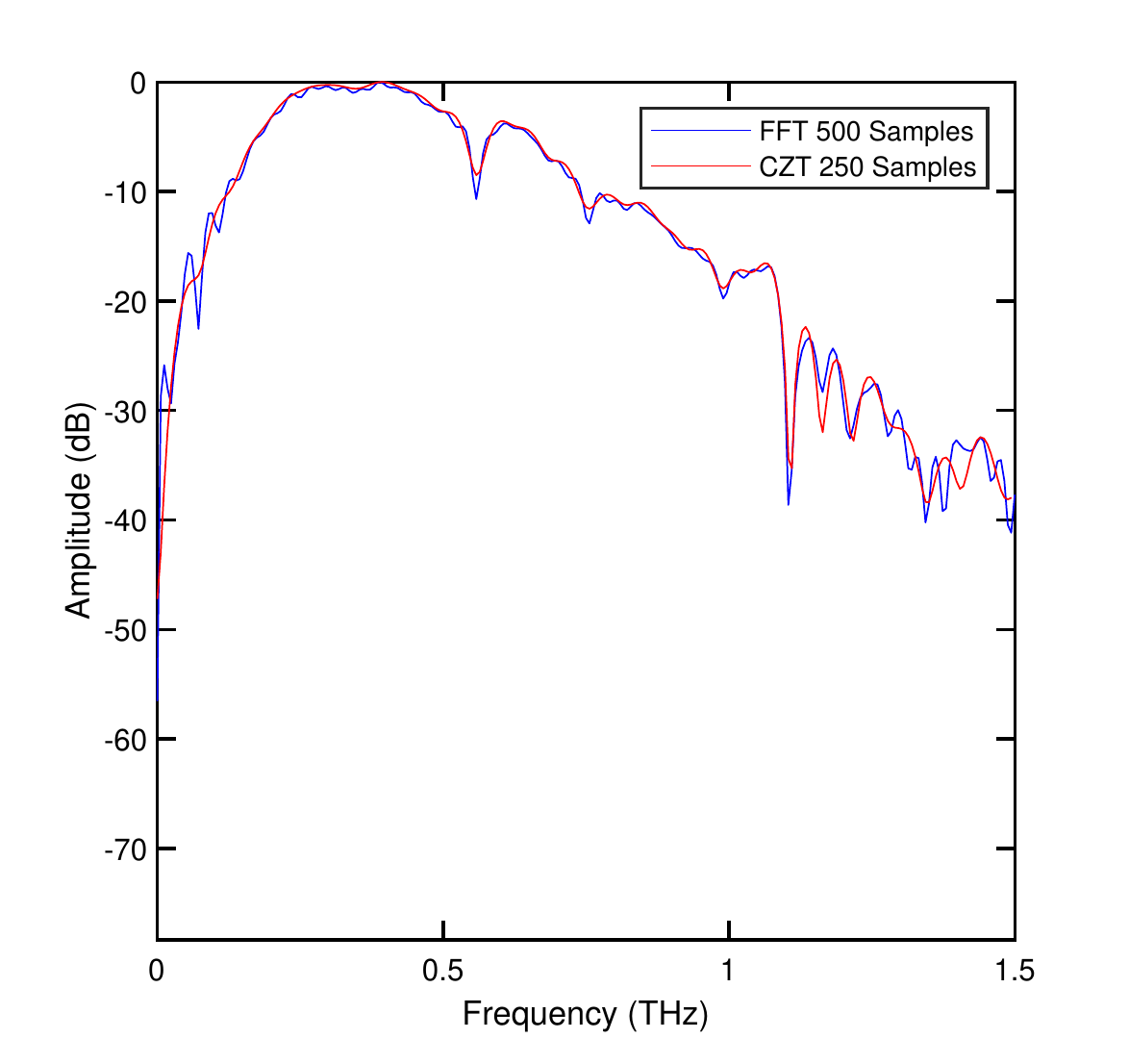}
\caption{Comparison of FFT with 500 samples and CZT with 250 samples.}
\label{fig_comp_3}
\end{figure}

A comparison of CZT with 250 samples and FFT with 500 samples at the same resolution gives Figure \ref{fig_comp_3}.

To achieve the same resolution it has been necessary to zero pad in Figure \ref{fig_comp_3}. It can be observed that with fewer samples, the CZT is not able to show the same information even if they have the same plotting resolution so it does not improve the frequency resolution.

\section{Computation perfomance comparison}
The time efficiency of each method has been evaluated for THz-TDS signals. Tests have been performed in which each method is run 10000 times for each spectral resolution and the average time it takes to run is calculated using MATLAB. Each method has been referenced to the time taken for the first operation, corresponding to a resolution of 20 GHz. 

\begin{figure}[htbp]
\centering
\includegraphics[width=6.5cm]{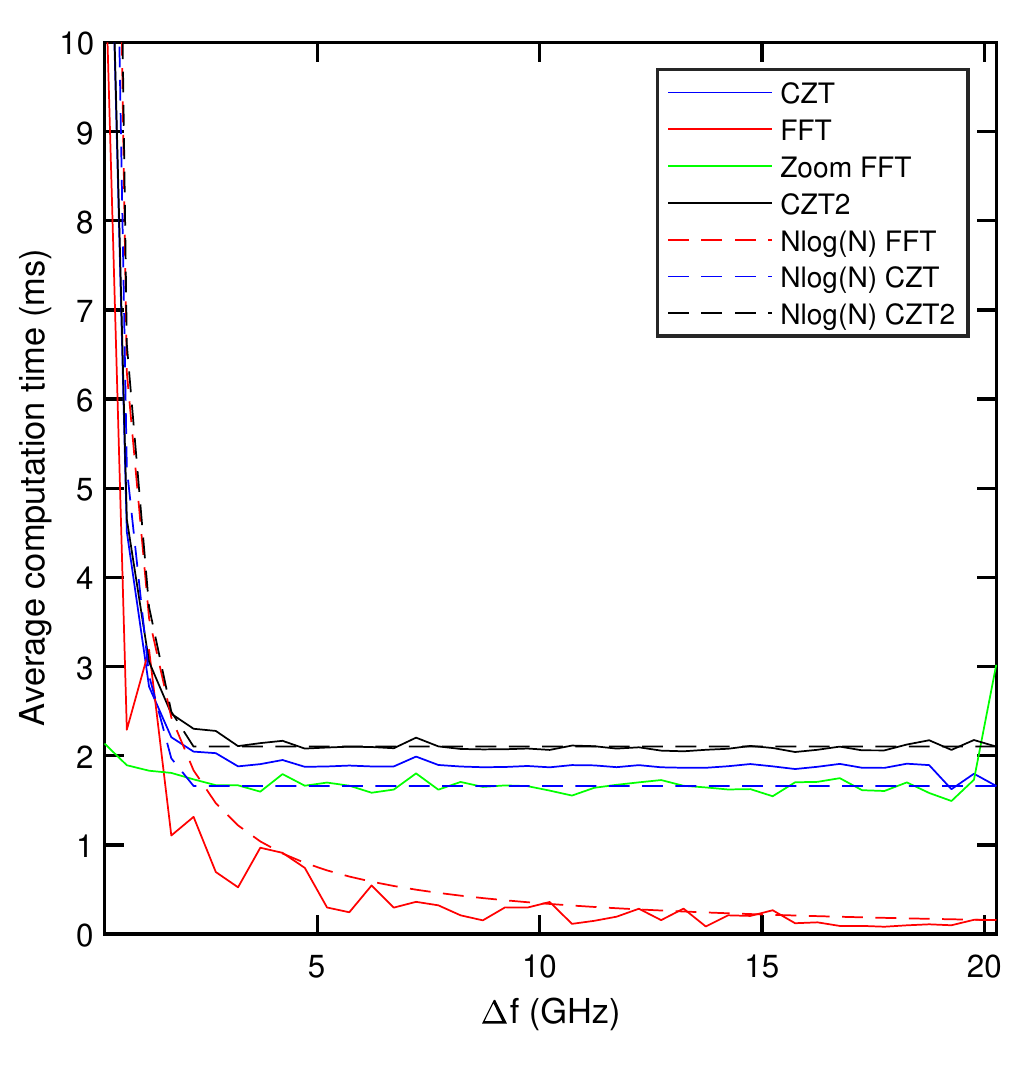}
\caption{Comparison of the execution time of the different methods with respect to the resolution at 0.5 GHz step. Solid: Computation time for FFT, Zoom FFT and CZT. Dashed: Theoretical cost.} 
\label{fig_time}
\end{figure}

In Figure \ref{fig_time} it can be seen that FFT is significantly faster than CZT. Zoom FFT behaves linearly in the high resolution (small $\Delta f$) range since the number of samples required to increase resolution changes very slightly. However, all the other methods need increasingly many samples in order to improve resolution. Zoom FFT only improves on FFT by adding about 10 times more zeros than samples. Looking at the CZT in \cite{b16} (CZT2), it can be seen that it is not as efficient as the CZT implemented in MATLAB.

\begin{figure}[H]
\centering
\includegraphics[width=6.5cm]{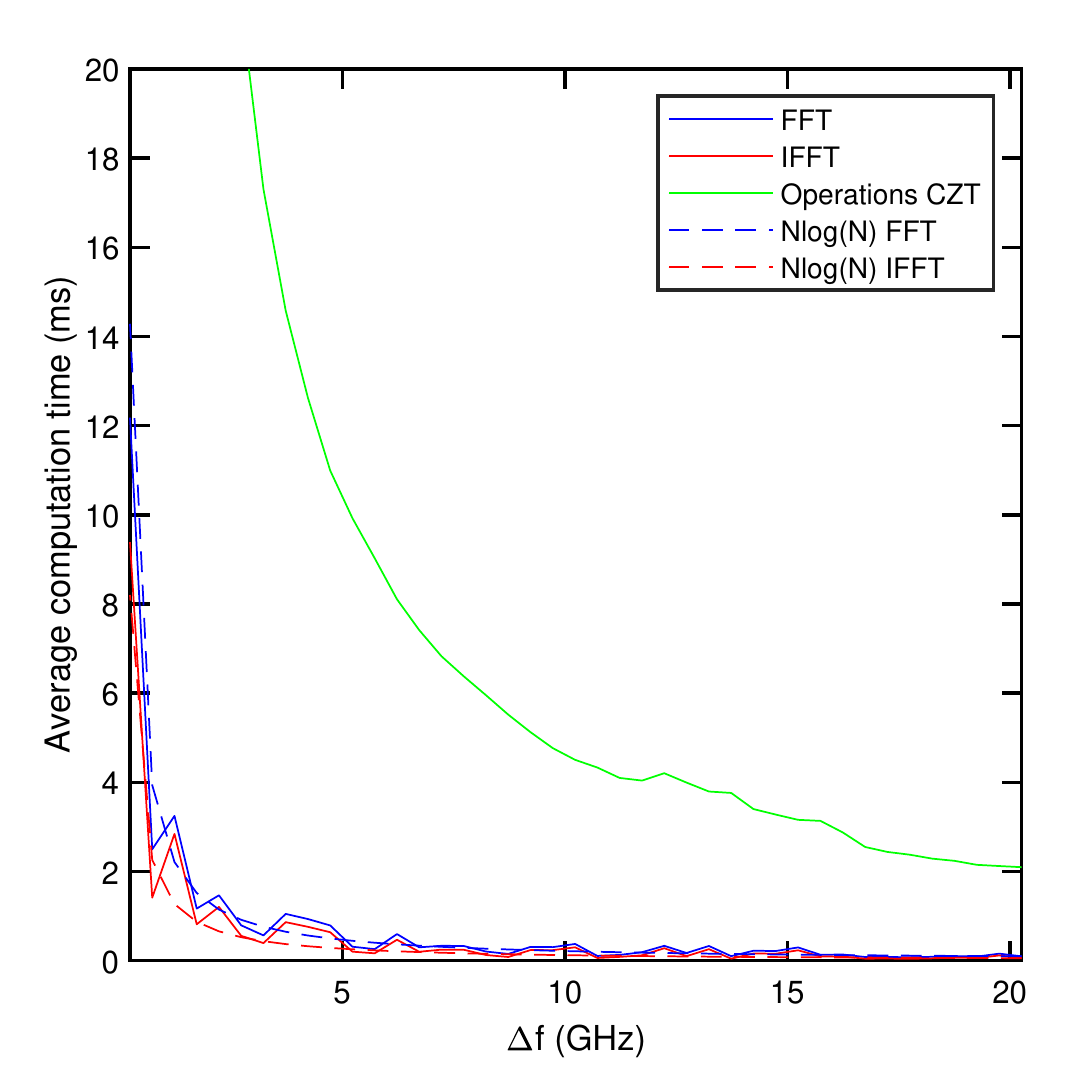}
\caption{Comparison of FFT and IFFT run time versus resolution at 0.5 GHz step.} 
\label{fig_fftvsifft}
\end{figure}

As CZT uses FFT twice and IFFT once, it should have a time cost relative to the sum of these three operations. It can be seen that there is a large difference in cost between CZT and FFT. This difference could be caused by the cost of the IFFT, but as seen in Figure \ref{fig_fftvsifft}, the IFFT has the same cost as the FFT, so the cost of the CZT is not only due to the cost of the sum of two FFTs and one IFFT, but also to the cost of the four complementary operations that are performed prior to the other operations. This difference may also be caused by the strong optimisation of the FFT in MATLAB through parallelisation of the code.

\section{Conclusion}
From simulations and experimental data, it has been show that the FFT method is the most time-efficient method of performing a frequency transformation in THz. The CZT and Zoom FFT manage to reduce the number of points compared to the FFT, but at no point do they surpass it in time efficiency. These transforms could have different applications where large data sizes are required, such as 2D or 3D transforms where the number of points grows with $N^2$ or $N^3$. Regarding resolution, a distinction can be made between plotting resolution and frequency resolution. The plotting resolution improves in the CZT and Zoom FFT transforms, as we restrict the frequency area where we concentrate the points, so with the same number of points the same resolution is achieved. The frequency resolution never improves because we do not manage to extract more information from the signal.

\section{Acknowledgements}
This work was supported in part by project PID2019-111339GB-I00 Spanish Ministerio de Ciencia, Innovación y Universidades-Agencia Estatal de Investigación.


\end{document}